\documentclass[useAMS,usenatbib]{mn2e}

\usepackage{graphicx}
\usepackage{url}
\usepackage{bm}

\newcommand{\sassy}{{\rm SASSy}}
\newcommand{\oracdr}{{\sc orac-dr}}
\newcommand{\smurf}{{\sc smurf}}

\newcommand{\hii}{H\,{\sc {ii}}}

\title{A Pilot Study for the SCUBA-2 `All-Sky' Survey}

\author[T. MacKenzie et al.]{Todd Mackenzie,$^{1}$\thanks{E-mail: todd@phas.ubc.ca}
Filiberto G. Braglia,$^{1}$ 
Andy G. Gibb,$^{1}$ 
Douglas Scott,$^{1}$\newauthor
Tim Jenness,$^{2}$
Stephen Serjeant,$^{3}$
Mark Thompson,$^{4}$
David Berry,$^{2}$\newauthor
Christopher M. Brunt,$^{5}$
Edward Chapin,$^{1}$
Antonio Chrysostomou,$^{2}$\newauthor
Dave Clements,$^{6}$
Kristen Coppin,$^{7,8}$
Frossie Economou,$^{2}$
A. Evans,$^{9}$
Per Friberg,$^{2}$\newauthor
Jane Greaves,$^{10}$
T. Hill,$^{11}$
Wayne Holland,$^{12,13}$
R. J. Ivison,$^{12,13}$
Johan H. Knapen,$^{14,15}$\newauthor
Neal Jackson,$^{16}$
Gilles Joncas,$^{17}$
Larry Morgan,$^{18}$
Chris Pearson,$^{19,20,3}$\newauthor
Michele Pestalozzi,$^{21}$
Alexandra Pope,$^{22}$
John Richer,$^{23,24}$
J. S. Urquhart,$^{25}$\newauthor
Mattia Vaccari,$^{26}$
Bernd Weferling,$^{27}$
Glenn White,$^{3,19}$
Ming Zhu$^{28}$
\\
$^{1}$Department of Physics \& Astronomy, University of British Columbia, 6224 Agricultural Road, Vancouver, BC V6T 1Z1, Canada\\
$^{2}$Joint Astronomy Centre, 660 North A'ohoku Place, Hilo, HI 96720, USA\\
$^{3}$Department of Physics \& Astronomy, The Open University, UK\\
$^{4}$Centre for Astrophysics Research, Science \& Technology Research Institute, University of Hertfordshire, College Lane, Hatfield, AL10\\ 9AB, UK\\
$^{5}$School of Physics, Stocker Road, Exeter, EX4 4QL, UK\\
$^{6}$Imperial College London, South Kensington Campus, London SW7 2AZ, UK\\
$^{7}$Department of Physics, McGill University, 3600 rue University, Montr{\'e}al, QC H3A 2T8, Canada\\
$^{8}$Institute for Computational Cosmology, Durham University, South Road, Durham DH1 3LE, UK\\
$^{9}$Astrophysics Group, Keele University, Keele, Staffordshire, ST5 5BG, UK\\
$^{10}$School of Physics \& Astronomy, University of St Andrews, North Haugh, St Andrews KY16 9SS, Scotland\\
$^{11}$Laboratoire AIM, CEA/IRFU -- CNRS/INSU -- Universit{\'e} Paris Diderot, CEA-Saclay, 91191 Gif-sur-Yvette Cedex,
France\\
$^{12}$UK Astronomy Technology Centre, Royal Observatory, Blackford Hill, Edinburgh EH9 3HJ, UK\\
$^{13}$Institute for Astronomy, University of Edinburgh, Blackford Hill, Edinburgh EH9 3HJ, UK\\
$^{14}$Instituto de Astrof{\'\i}sica de Canarias E-38200 La Laguna, Tenerife, Spain\\
$^{15}$Departamento de Astrof{\'\i}sica, Universidad de La Laguna, E-38205 La Laguna, Tenerife, Spain\\
$^{16}$University of Manchester, School of Physics \& Astronomy, Jodrell Bank Centre for Astrophysics,
Alan Turing Building, Oxford Road,\\ Manchester, M13 9PL, UK\\
$^{17}$D{\'e}pt. de physique, de g{\' e}nie physique et d'optique and Centre de recherche en astrophysique du Qu{\' e}bec,
Universit{\' e} Laval, Qu{\' e}bec,\\ G1V 0A6, Canada\\
$^{18}$Astrophysics Research Institute, Liverpool John Moores University, Twelve Quays House, Egerton Wharf, Birkenhead, Wirral,\\ CH41 1LD, UK\\
$^{19}$RAL Space, Rutherford Appleton Laboratory, Chilton, Didcot, Oxfordshire OX11 0QX, UK\\
$^{20}$Institute for Space Imaging Science, University of Lethbridge, Lethbridge, Alberta T1K 3M4, Canada\\
$^{21}$IFSI/INAF, via del Fosso del Cavaliere 100, I--00173 Roma, Italy\\
$^{22}$Department of Astronomy, University of Massachusetts, 710 North Pleasant Street, Amherst, MA 01003-9305,
USA\\
$^{23}$Astrophysics Group, Cavendish Laboratory, J. J. Thomson Avenue, Cambridge CB3 0HE, UK\\
$^{24}$Kavli Institute for Cosmology, c/o Institute of Astronomy, University of Cambridge, Madingley Road,
Cambridge CB3 0HA, UK\\
$^{25}$Australia Telescope National Facility, CSIRO Astronomy and Space Science, Sydney, NSW 2052, Australia\\
$^{26}$Department of Astronomy, University of Padova, Vicolo Osservatorio 3, I-35122, Padova, Italy\\
$^{27}$University Bamberg, Markusplatz 3, 96045, Bamberg, Germany\\
$^{28}$National Astronomical Observatory of China, Beijing, China\\
\vspace{3in}
}

\begin{document}

\maketitle{}

\begin{abstract}
We have carried out a pilot study for the SCUBA-2 `All-Sky' Survey, SASSy, a wide and shallow
mapping project at $850\,\mu$m, designed to find rare objects, both Galactic and extragalactic.
Two distinct sets of exploratory observations were undertaken, and used to test the SASSy approach and
data reduction pipeline.  The first was a $0.5^\circ\times0.5^\circ$ map around the nearby galaxy
NGC 2559.  The galaxy was easily detected at $156\,$mJy, but no other convincing sources are present
in the map.  Comparison with other galaxies with similar wavelength coverage indicates that NGC 2559
has relatively warm dust.  The second observations cover $1\,{\rm deg}^2$ around the W5-E \hii\ region.
As well as diffuse structure in the map, a filtering approach was able to extract
27 compact sources with signal-to-noise greater than 6.  By matching with data at other wavelengths we
can see that the SCUBA-2 data can be used to discriminate the colder cores.  Together these observations
show that the SASSy project will be able to meet its original goals of detecting new bright sources
which will be ideal for follow-up observations with other facilities.
\end{abstract}


\section{Introduction}\label{sec:intro}

The millimetre and sub-millimetre parts of the electromagnetic spectrum directly probe
the cold Universe.  The sub-mm window specifically allows us to study the youngest phases 
of star formation in our Galaxy, and the dusty, most prodigiously star-forming galaxies at high redshift.
Despite this strong motivation, the sub-mm sky still remains poorly surveyed.
The SCUBA-2 `All-Sky' Survey, or \sassy\footnote{Alternatively known as
the SCUBA-2 Ambitious Sky Survey.}, is a James Clerk Maxwell Telescope (JCMT) Legacy Survey (JLS)
project designed to redress this 
balance and exploit the rapid mapping capability of SCUBA-2 to ultimately map a large portion of 
the sky visible from the JCMT to an angular resolution of 14 arcsec at $850\,\mu$m.  The target
point source rms level is $30\,$mJy.

The benefits of 
such a wide-field survey are many, ranging from a complete census of infrared dark clouds (IRDCs) 
to the potential discovery of some of the most luminous high-redshift galaxies in the Universe \citep{sassy}.
The approved phase of SASSy consists of two distinct parts:
a strip covering the Galactic Plane which is visible from Hawaii; 
and a `Pole-to-Pole' strip perpendicular to this and designed to pass through the Galactic 
and Ecliptic North Poles.  These observations will be carried out in `Grade~4' weather conditions 
($0.12<\tau_{225}<0.2$, where $\tau_{225}$ is the sky opacity at $225\,$GHz, as measured by the
CSO radiometer), i.e.\ essentially when the 
atmosphere is too opaque to enable useful observations 
of fainter objects with SCUBA-2.  The $450\,\mu$m data are therefore expected to be of marginal value, and 
the survey is entirely designed to make large maps at $850\,\mu$m, with a
target sensitivity of $30$\,mJy, achieved using a fast scanning speed.

SASSy will be able to build on the success of {\it IRAS\/} at one decade shorter wavelengths.  It is
also complementary to several more recent wide surveys, namely {\it WISE\/} and {\it Akari}, at near-
and mid-IR wavelengths, and surveys with the {\it Herschel\/} and {\it Planck\/} satellites in the submillimetre
(hereafter submm).  Early results
from {\it Herschel\/} have already demonstrated that bright lensed galaxies can be selected at submm
wavelengths \cite{Negrello}, and also that potentially protostellar cores can also be found using {\it Herschel\/}
\cite{Derek}.  The {\it Planck\/} Early Release Compact Source Catalogue will contain an all-sky list of sources
which includes the $850\,\mu$m channel.  However, the resolution of SASSy will be 20 times better.  Hence it is
clear that if SCUBA-2 is able to map rapidly enough while maintaining its nominal beamsize and achieving the
required sensitivity, then SASSy will be able to meet its science goals.  As we show below, these preliminary
SCUBA-2 data lend confidence that these requirements can be met and that SASSy will discover many new and
interesting sources.

In the next section we describe the two sets of observations which were made as a pilot study for SASSy, one
Galactic field and one extragalactic field.  In Section~\ref{sec:reduction} we describe how we reduced the data
and extracted sources.  We then discuss the properties of the sources that we found and end with some
conclusions about SASSy in general.

\section{Observations}\label{sec:observations}

SCUBA-2 \citep{2003scuba2, 2006scuba2} is the successor to the Submillimetre Common 
User Bolometer Array (SCUBA, \citealt{1999scuba}), which operated successfully on the JCMT from 1997--2005.  
SCUBA-2 has been designed to be hundreds of time faster
than SCUBA for mapping the sky in the same two primary bands, $450\,\mu$m and $850\,\mu$m.
The observatory offered a period of `Shared Risks Observing' programme (hereafter S2SRO), in which 
relatively short programmes were carried out with a partially-commissioned version of 
the instrument, having one (of four) sub-array available at each of the two bands.  
Unless stated otherwise, all maps and detections in this paper are at $850\,\mu$m,
since this is the primary wavelength of interest for SASSy.

\begin{figure}
  \centering
  \includegraphics[width=3.25in]{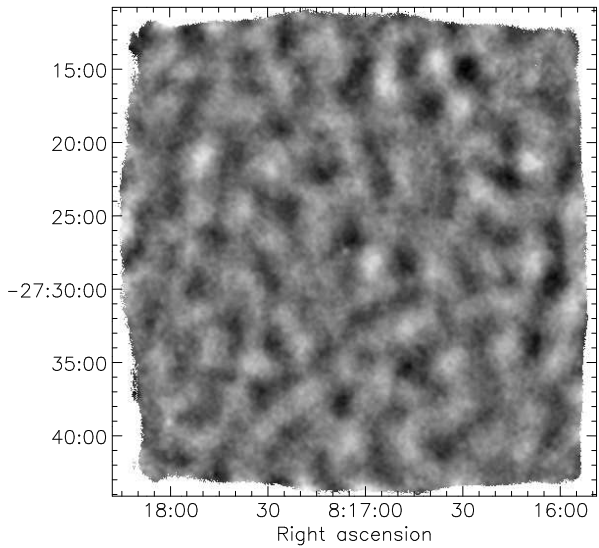}
  \caption{Smoothed signal-to-noise map of the `extragalactic' field, being about $0.5^\circ$ across.
NGC 2559 is located at the centre of this map, which is dominated by artefacts of
roughly the SCUBA-2 array size.}\label{smoothedngc2559}
\end{figure}
\begin{figure}
  \centering
  \includegraphics[width=3.25in]{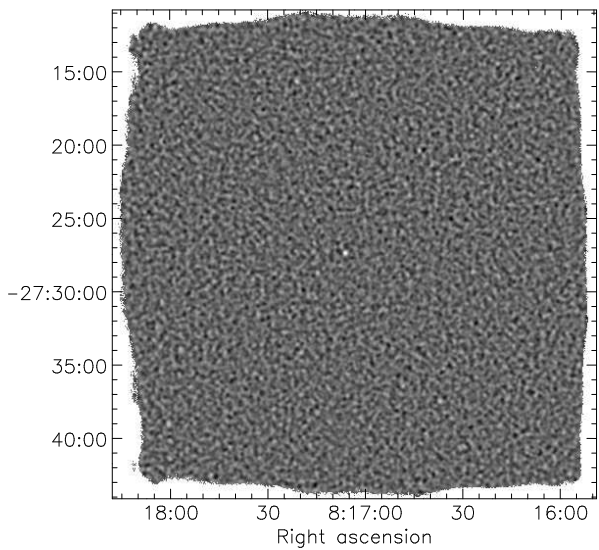}
  \caption{Matched-filtered signal-to-noise
map of the `extragalactic' field. NGC 2559 is located at the centre of this map, and
is clearly detected.  Other peaks in this map (near the north-east and south-east corners) appear to be just noise excursions.}\label{mfngc2559}
\end{figure}

As part of S2SRO there were two separate sets of observations carried out for SASSy, which we now describe.
The SCUBA-2 array of $32\times40$ transition-edge sensitive bolometers has a footprint of about
3 arcmin square on the sky (it will double in linear size when the focal plane is fully populated with
sub-arrays) on the sky, and takes data at a sampling rate of approximately $200\,$Hz while scanning across the selected region. The
available $850\,\mu$m array typically only had 50 per cent of the detectors working (although the number
used in the data reduction is selected dynamically and hence varies with time).  Calibration was performed
using an internal flat-field source, as well as absolute measurements using known calibrators on the sky.

\subsection{Extragalactic observations}

The first set of observations planned was imaging of 
a region measuring $0.5^\circ\times 2^\circ$, with
one end centred on NGC\,2559, a nearby galaxy, and the other end
extending into the Galactic Plane cirrus.  NGC\,2559 is a spiral galaxy at a distance
of 20.8\,Mpc (from the systemic velocity and assuming a Hubble constant of 75\,km\,s$^{-1}$),
with morphological type SB(s)bc pec \citep{deV91}.
It was chosen as a target in this 
study since it is the IR-brightest nearby galaxy that lies in the SASSy area, and
had not been previously observed by SCUBA.
This section of sky was broken up into four $0.5^\circ\times 0.5^\circ$
tiles, of which only two were observed during the S2SRO period -- these were the
outer ends of the strip, and hence are not contiguous.
We refer to this as the `extragalactic' pilot survey, 
since (although relatively close to the Galactic Plane) the field was chosen to 
contain a bright {\it IRAS\/} galaxy which had not previously been observed at submillimetre wavelengths.

The region containing NGC\,2559 was observed on 2010 February 27 and March 14,
for a total of 167 minutes of observing yielding an average integration time of 27 seconds per pixel.
The average optical depths at $225$\,GHz for the two nights were $\tau_{225}=0.097$ and 0.154,
the average noise-equivalent flux densities were 160 and 260 mJy$\,{\rm s}^{-1/2}$,
and the telescope scan rates were 240 and 360 arcsec\,${\rm s}^{-1}$, respectively.
Data were reduced using the
\smurf\ map-maker (Sub-Millimetre User Reduction Facility, \citealt{smurf,smurfADASS}, Chapin et
al.~in preparation), which we describe in more detail in the next section.  The resulting raw
map has a noise of $38\,{\rm mJy}\,{\rm beam}^{-1}$ determined using the produced noise map.

The region about $1^\circ$ away, containing known Galactic cirrus,
was observed for a total of 105 minutes, resulting in an average integration time of 16 seconds per pixel.  
Both `pong' and `rotating pong' scan strategies \citep{scanpatterns}\footnote{`Pong'
is the default scanning mode for covering large areas with SCUBA-2, and is like a raster-scanning
pattern except that it is designed to visit different parts of the map on different timescales.  `Rotating
pong' means that the orientation of the pattern is allowed to rotate in sky coordinates as the observations
are carried out, resulting in scans at many different angles, which is better from a map-making perspective.}
were used for the NGC\,2559 field and only the rotating pong strategy for the cirrus field.
Although NGC\,2559 is readily detected in the resulting map of this area,
we are unable to identify any Galactic cirrus
within the other map.  This is due to a combination
of the relatively high noise level, and the difficulty in detecting extended diffuse structure
(largely removed in the reduction process).
Therefore for the remainder of this paper, of the two extragalactic observations, 
we restrict our attention to the map containing the galaxy
NGC\, 2559.

\subsection{Galactic Observations}

The second distinct set of observations targetted
a field around the W5-E \hii\ region.  This region was selected because of its simple geometry and, at 2\,kpc \citep[e.g.][]{karr}, is one of the
nearest regions of triggered massive star formation \citep{2008hsf1.book..264M}.  A single region of $1^\circ\times1^\circ$ was mapped on 2009 December 5, 
at a speed of 600 arcsec\,s$^{-1}$ (the nominal scan-rate for \sassy) using the
`pong' scanning mode \citep{scanpatterns}.  The scan pattern had each successive sweep separated by 120 arcsec,
ensuring an overlap to improve mapping performance.

The on-source time was 70 minutes which resulted in a total of 12 passes over
the entire region. In addition the central $0.5^\circ\times 0.5^\circ$
was mapped at the same speed for another 70 minutes, covering that region 42 times,
for a total observing time of 140 minutes.  The average integration time per pixel for the inner and outer
regions were 18 and 3 seconds respectively.
The 225-GHz optical depth varied from 0.08 to 0.12 with a mean of 0.10
(corresponding to Grade 3 weather). This is substantially lower than
the allotted opacity band for \sassy; however, the central $0.25\,{\rm deg}^2$
region of the map has a sensitivity slightly exceeding
the \sassy\ target level, and therefore represents a valid test of
the detectability of sources. The central portion of the map has a noise
of $25\,{\rm mJy}\,{\rm beam}^{-1}$; the value for the outer region is $60\,{\rm mJy}\,{\rm beam}^{-1}$, as reported by the noise map.
Note that $450\,\mu$m data were also obtained and were also reduced, but were found to be of limited use, with
only the single brightest source being detected.

\section{Data Reduction and Source Detection}\label{sec:reduction}

The raw timeseries data were processed using \smurf\ \citep{smurf,smurfADASS} called from the
\oracdr\ pipeline \citep{oracdr}. \smurf\ solves for the astronomical
signal using an iterative technique, fitting and filtering out noise contributions
from the atmosphere and the instrument. Details of the map-maker may
be found in Chapin et al.\ (2010, in preparation). Readings from the JCMT water-vapour
monitor (WVM, \citealt{wvm}) were used to correct for atmospheric extinction.
The \oracdr\ pipeline was used to mosaic the individual observations using
inverse-variance weighting.

\begin{figure}
  \centering
  \includegraphics[width=3.5in]{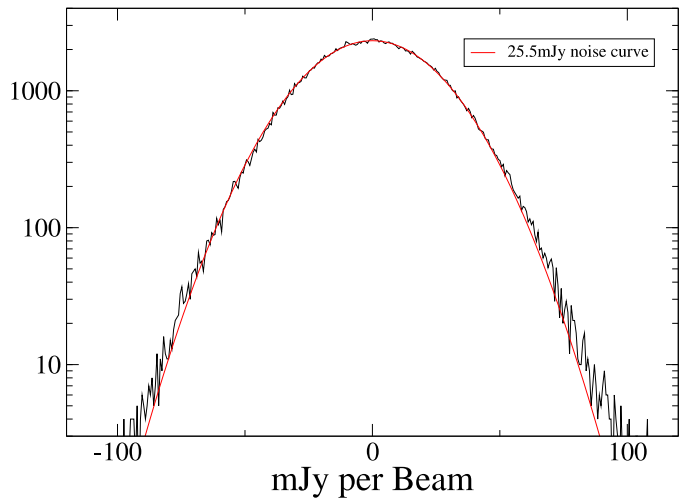}
  \caption{Histogram of matched-filtered pixel values in the `extragalactic' map.
This can be used to determine the spatial noise in the map.  A $25.5\,$mJy noise curve
is plotted for comparison, while the expected uncorrelated noise in the map is significantly smaller.}\label{noisehistogram}
\end{figure}

In the NGC 2559 raw map it is clear that time-dependent noise shows up as
large-scale structure, obscuring the detection of the galaxy.  This comes from a combination of
residual sky fluctuations, as well as oscillations inherent to the instrument when these data were taken.
Fig.~\ref{smoothedngc2559} shows the NGC 2559 field after smoothing with the beam 
(to improve the signal-to-noise ratio for sources).  The resulting map is dominated by 
structure at roughly the scale of the SCUBA-2 array.  However, since the core of the 
galaxy shows up as a compact source, we are able to use a point source filter
to make an unambiguous detection.  The `matched-filter' method implemented
within \oracdr\ subtracts the map smoothed with a larger Gaussian (30 arcsec in this case)
from the map convolved with a Gaussian equal to the JCMT beamsize (14 arcsec
at $850\,\mu$m), thus giving a `Mexican hat' type spatial filter.
The resulting map is able to enhance sources that 
are approximately point-like (since we expect extra-galactic sources found by SASSy to be approximately point-like).  
We experimented with different choices for filter shape, and 
found that the default within \oracdr\ is close to the best we can do in terms of signal-to-noise ratio.  
In principle we could improve things by using specific
knowledge of the expected shape of NGC 2559, but that would not be helpful
in a blind SASSy search (which is what we are preparing here).

Fig.~\ref{mfngc2559} shows the NGC 2559 field after applying the matched-filter algorithm
to remove the large-scale structure.  NGC 2559 is now plainly visible in the 
centre of the map.  Fig.~\ref{noisehistogram} shows a histogram of the pixel values
found in the matched-filtered map of NGC 2559.  After removing the background variations in the map, 
the spatial noise is found to be approximately $25.5\,{\rm mJy}$ (equivalent for a point source).
This is larger than the average noise of $19\,{\rm mJy}$ calculated by the data-reduction pipeline
(i.e.\ given by the noise map),
due to not having filtered out all the spatially correlated noise.
Since shallow extragalactic fields should be composed of white noise plus a few sources, the signal-to-noise
matched-filtered map is re-normalized to have an rms of unity before searching for sources.
In other words we are effectively using the spatial rms noise, rather than the purely white
noise estimate which comes from the pipeline noise map.

\begin{figure*}
  \centering
  \includegraphics[width=6.5in]{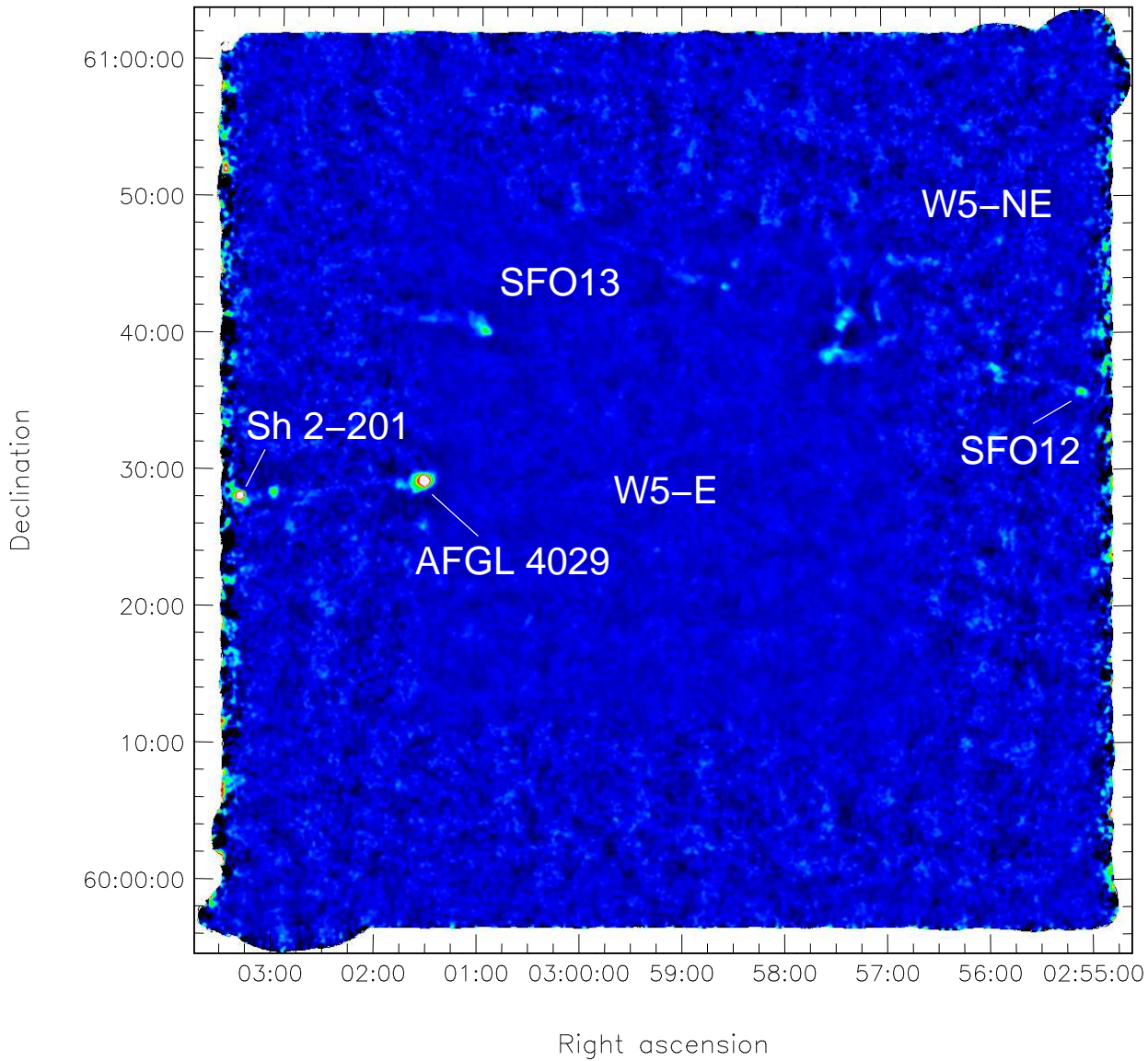}
  \caption{The W5-E star forming region as mapped by SCUBA-2 at $850\,\mu$m, smoothed with
a Gaussian with a FWHM of 14 arcsec. Colour scale ranges from $-100$ to $+500\,{\rm mJy}\,{\rm beam}^{-1}$.
The central roughly 1/4 of the map can be seen to have lower noise than the rest.
Known objects are labelled, as is the approximate centre of the W5-E \hii\ region.}\label{w5colsm2}
\end{figure*}

\begin{figure*}
  \centering
  \includegraphics[width=6.5in]{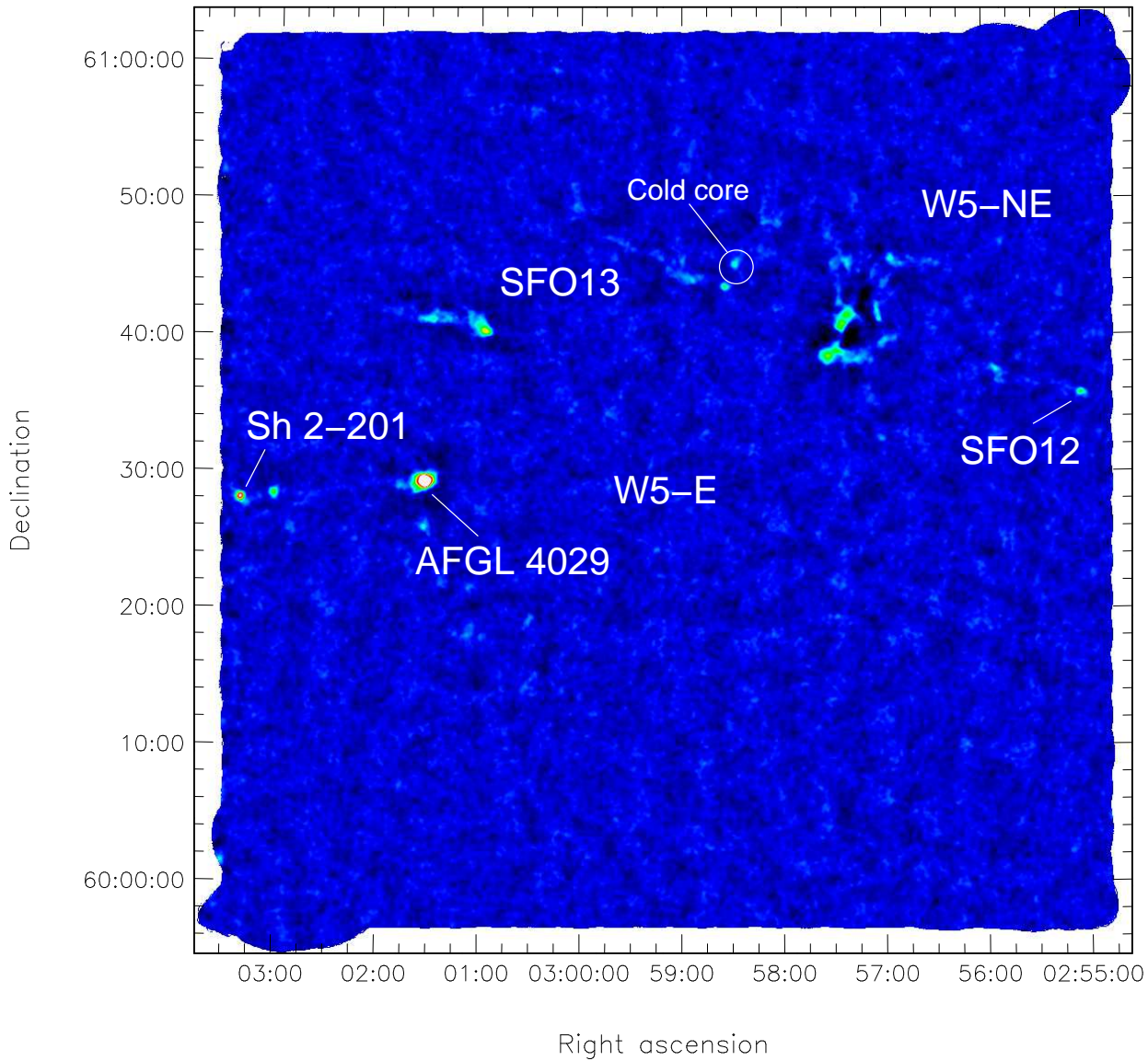}
  \caption{Signal-to-noise ratio image for W5-E. The colour scale ranges from $-2$ to $+10$. The `cold core' referred to in the text is circled.  Known objects are labelled, as is the approximate centre of the W5-E \hii\ region.}\label{w5colsnr}
\end{figure*}

\begin{figure}
  \centering
  \includegraphics[width=3.5in]{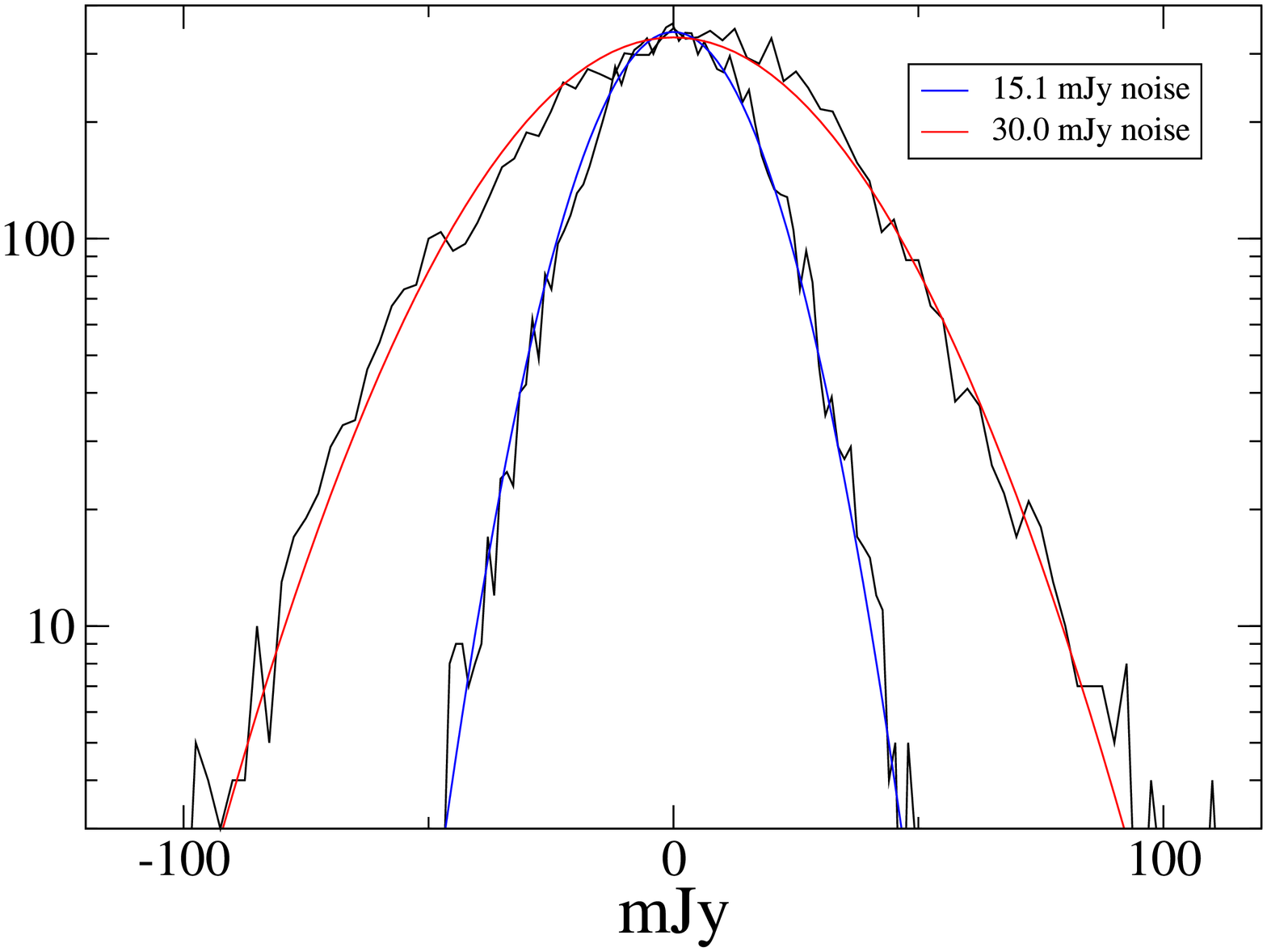}
  \caption{Histogram of matched-filtered pixel values in the W5-E map for 100 square pixel blank sections of the inner and outer regions of the map.
Noise curves were fit and determined to be 15 and $30\,{\rm mJy}$, respectively.}\label{w5noise}
\end{figure}

\begin{figure}
  \centering
  \includegraphics[width=3.25in]{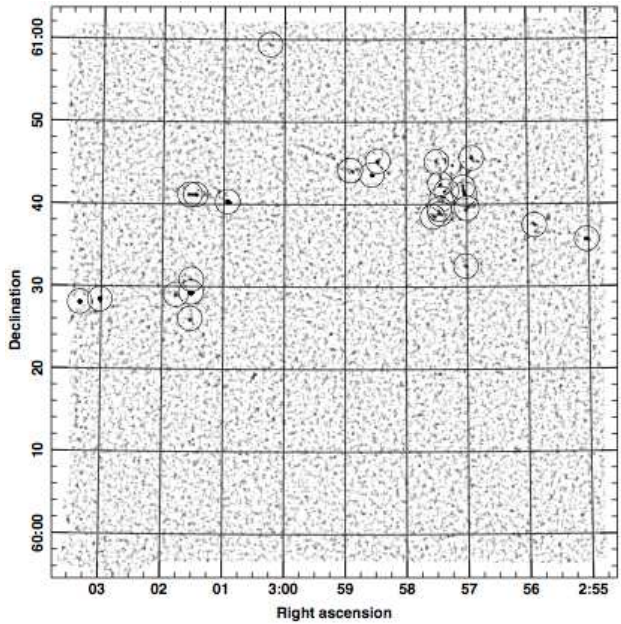}
  \caption{The matched-filtered map of W5-E with circles indicating the ${>}\,6\sigma$ sources listed in Table~\ref{w5objects}.}\label{w5mfsources}
\end{figure}

Fig.~\ref{w5colsm2} shows the W5-E Galactic star-forming region map, while Fig.~\ref{w5colsnr} shows the related
signal-to-noise map and Fig.~\ref{w5colsnr} shows an annotated version with several previously catalogued regions labelled.
Since the purpose of SASSy is to find and catalogue new sources for future study, it is simplest
to ignore source structure and to filter the map with the matched-filter.
For objects within the Galactic strip, we use a 60 arcsecond background subtraction for 
the matched-filter.  This scale is chosen since we expect Galactic sources to be more extended than extragalactic sources.
Fig.~\ref{w5mfsources} shows the W5-E map after being matched-filtered.
The central ${\sim}\,0.5^\circ\times0.5^\circ$ region has a spatial noise level of $15\,$mJy, while the outer
region has a noise level of approximately $30\,$mJy, once processed by the matched-filter (see Fig.~\ref{w5noise}).
Note that these values are different from the noise values in the unfiltered map since they only apply to the detection
of compact sources (and so should not be considered as being a noise `per beam', as would be usual for extended
structure).


\section{Detections}\label{sec:detections}

\subsection{NGC 2559 Field}

In order to extract sources from the matched-filtered maps in a manner which
could be easily automated for the full survey, 
we use the {\sc fellwalker} algorithm \citep{cupid}, 
implemented in the {\sc starlink} software package {\sc cupid}.  We use this algorithm
simply to associate contiguous blocks of pixels with a single
source -- the brightness is estimated from the peak value in the match-filtered signal-to-noise map.
Given the simple nature of the source detection, the choice of algorithm is not critical.
A minimum number of pixels of 7 and a low RMS level are
used in {\sc fellwalker} to recover as many real and noise peaks as possible for producing histograms.

\begin{figure}
  \centering
  \includegraphics[width=3.5in]{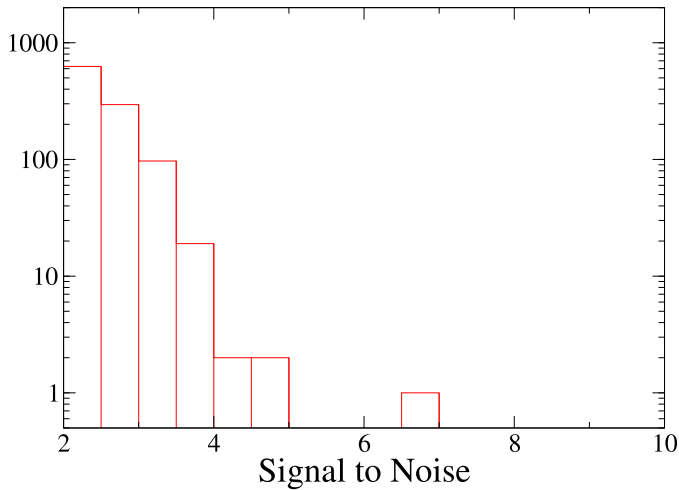}
  \caption{Histogram of peak signal-to-noise values within the matched-filtered NGC 2559 field.
NGC 2559 itself shows up at around $6.6\sigma$, while the 2 candidate sources at about $4.5\sigma$ appear to
be noise bumps (note that a $4\sigma$ event is not very unlikely, given the number of pixels in this map).}\label{clumphistogram}
\end{figure}

Fig.~\ref{clumphistogram} shows a histogram of signal to noise peaks of all the
sources within the NGC 2559 field.  NGC 2559 is clearly 
detected at a signal to noise of about 7, and there are two other candidate sources
at a signal to noise around 4.5. 
There are no obvious counterparts for these two sources in any relevant survey.  The histogram shows that
the detection of NGC 2559 is an outlier in the distribution, but the two other candidates seem consistent
with the noise distribution, and in a map with ${>}\,10^4$ beam-size pixels they are not very unlikely.
Nevertheless we checked for counterparts in the 
20-cm NVSS survey \cite{Condon98}.  We additionally
extracted archival 20-cm VLA data of approximately the same
depth (lower bandwidth but larger integration time), re-reduced them and
added them to the NVSS data.  No radio sources are detected in the combined 20-cm image
at the positions of the two peaks in the SASSy map,
where the noise level is 320--350$\,\mu$Jy.

We also explored different choices for the filtering, and found that although the detection of
NGC 2559 is robust, the next most significant peaks vary in position and brightness.
The SASSy survey plan
is to carry out short follow-up observations of such candidates
to distinguish between real objects and false positives (whether just noise excursions or
mapping artefects).  This pilot survey suggests that the level at which follow-up will
be worthwhile is around the $5\sigma$ level for large maps.

\begin{figure}
  \centering
  \includegraphics[width=3.25in]{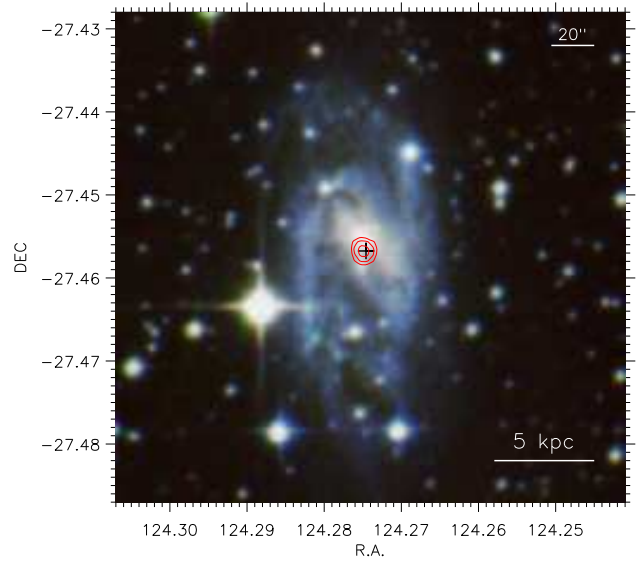}
  \caption{SCUBA-2 $850\,\mu$m 5 and $7\sigma$ contours
overlaid on an optical 3-colour image of NGC 2559 (derived from DSS data).
The black cross marks the position of the NVSS radio source.}\label{ngc2559new}
\end{figure}

Fig.~\ref{ngc2559new} shows an optical image (from the Digitized Sky Survey) of 
NGC 2559 with matched-filtered SCUBA-2 contours overlaid.
NGC 2559 shows up at a flux density of $156\pm26\,$mJy and would be detectable in data
representative of SASSy (with target rms of $30\,$mJy).


\subsection{W5-E Field}

Turning to the Galactic pilot map, this field contains a number of known sources, including three bright-rimmed clouds (SFO\,12, 13 and 14),
and the \hii\ region Sh~2-201. Note that SFO\,14 contains the massive young stellar object AFGL\,4029. The three bright-rimmed clouds have all been detected previously at submillimetre wavelengths using SCUBA \citep{morgan}.

\begin{figure}
  \centering
  \includegraphics[width=3.25in]{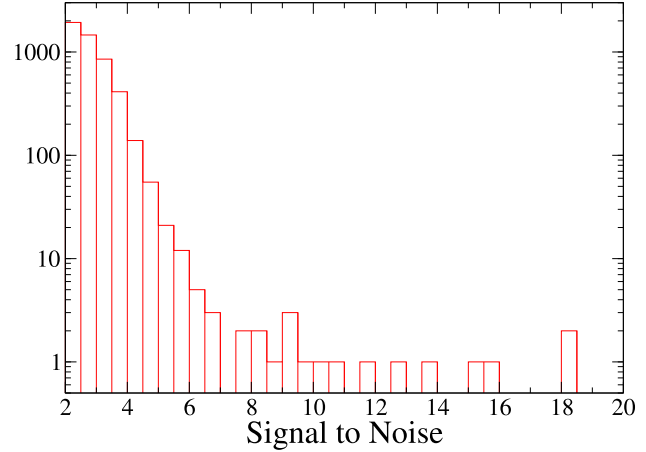}
  \caption{Histogram of peak signal-to-noise values within the matched-filtered W5-E field.  The brightest
 W5-E source (AFGL\,4029) has a signal-to-noise ratio of over 100 and is not shown.}\label{w5clumphistogram}
\end{figure}

All of these sources are detected in the SCUBA-2 map with good to high significance. A total of 27 sources were identified at $>$6-$\sigma$ after applying the matched-filter, most of which are unknown at submillimetre wavelengths. The central portion of the map containing the W5-E \hii\ region is devoid of dust emission. Table~\ref{w5objects} lists the objects found with a signal-to-noise ratio greater than 6 using the matched-filter method.  Of the objects in Table~\ref{w5objects}, 11 are brighter than $150$\,mJy and would be detected by a blind SASSy survey at more than $5\sigma$. Fig.~\ref{w5clumphistogram} shows a histogram of signal-to-noise peaks of all the sources extracted from the W5-E map.

By inspecting the unfiltered map it is clear that some of the `sources' found in this way are parts of extended filamentary structures. Fig\,\ref{w5colsm2} shows a number of extended, filamentary features of low signal-to-noise ratio. While faint, they are undoubtedly real as they show good agreement with the CO data presented by \citet{karr} and \citet{niwa}, as well as {\it Spitzer} MIPS images \citep{2008ApJ...688.1142K}. However, in practice, mapping such low surface-brightness features is beyond the scope of \sassy\ and falls into the realm of followup observations triggered by detecting new compact sources. The most interesting Galactic sources found by SASSy will be relatively isolated, and they will be discovered through applying a simple automated source-extraction procedure similar to that used here. Nonetheless it is encouraging to find that \sassy\ is capable of detecting extended features and there are many approaches which can be taken to characterise such morphology. Further experience with \sassy\ data will guide our approach.

\begin{table}
\centering
\begin{tabular}{rlcc} \hline
\multispan2 $S_{850}$ & \multispan2 \hfil Position (J2000)\hfil\\
mJy & S/N & RA & Dec \\ \hline
2290 & 163 & 03:01:31.783 & +60:29:19.81 \\
1970 & 44.2 & 03:03:21.039 & +60:28:03.98 \\
410 & 16.0 & 02:55:01.806 & +60:35:43.70 \\
405 & 18.0 & 03:03:01.319 & +60:28:24.23 \\
236 & 18.1 & 03:00:56.349 & +60:40:20.11 \\
226 & 9.0 & 02:56:54.974 & +60:45:37.30 \\
212 & 8.6 & 02:55:53.335 & +60:37:29.14 \\
185 & 15.2 & 02:58:33.795 & +60:43:36.67 \\
172 & 12.6 & 02:57:02.215 & +60:41:15.56 \\
164 & 6.7 & 03:01:46.608 & +60:29:03.05 \\
152 & 13.8 & 02:57:24.688 & +60:40:40.34 \\
131 & 11.6 & 02:57:33.643 & +60:38:28.77 \\
130 & 6.7 & 03:00:14.481 & +60:59:22.56 \\
119 & 9.1 & 03:01:32.609 & +60:26:08.48 \\
118 & 10.5 & 02:58:27.730 & +60:45:15.82 \\
112 & 9.6 & 02:57:26.183 & +60:38:51.92 \\
110 & 8.4 & 02:57:03.894 & +60:42:04.86 \\
108 & 10.1 & 02:57:19.897 & +60:41:37.63 \\
99 & 8.2 & 02:57:26.221 & +60:42:31.99 \\
89 & 6.2 & 03:01:31.556 & +60:30:53.96 \\
87 & 7.7 & 02:57:27.009 & +60:39:26.70 \\
79 & 6.0 & 02:58:55.106 & +60:44:08.31 \\
77 & 7.7 & 03:01:27.812 & +60:41:10.03 \\
75 & 6.6 & 02:57:01.063 & +60:32:31.05 \\
71 & 6.1 & 02:56:59.793 & +60:39:31.14 \\
69 & 6.2 & 02:57:29.653 & +60:45:10.17 \\
68 & 6.2 & 03:01:33.497 & +60:41:10.29 \\
\end{tabular}
\caption{List of objects found with a peak signal-to-noise greater than 6 in the W5-E region
using the matched-filter method.
}\label{w5objects}
\end{table}

\begin{figure}
  \centering
  \includegraphics[width=3.25in]{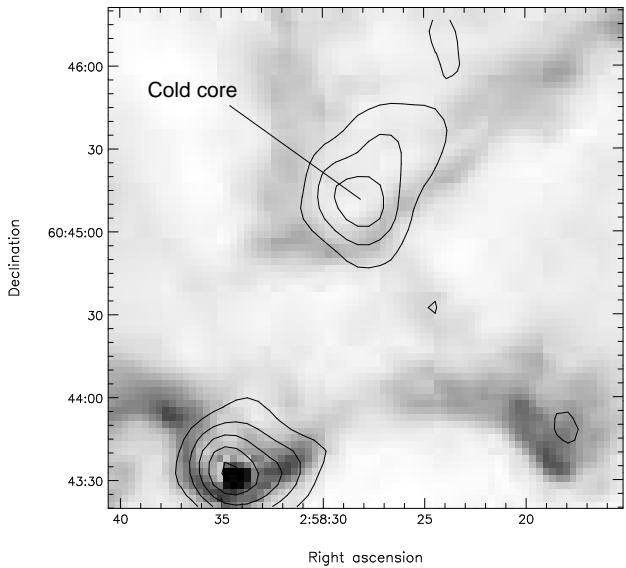}
\caption{SCUBA-2 contours overlaid on {\it Spitzer\/} 24\,$\mu$m image. The source labelled `Cold core' is circled
and shows up clearly in the submm, yet within a dark region of the {\it Spitzer\/} image. It is, however,
detected at 70 and 160\,$\mu$m with {\it Spitzer}.}\label{coldcore}
\end{figure}

Of all the new submillimetre detections perhaps the most striking source in the SCUBA-2 map is that labelled as the `cold core' in Fig.~\ref{w5colsnr}.
It shows up clearly in the signal-to-noise ratio map at a significance of 7$\sigma$. Comparison with {\it Spitzer\/}
data reveals that while it is also detected at 160 and 70\,$\mu$m, it is completely absent at 24\,$\mu$m.
As shown in Fig.~\ref{coldcore}, the SCUBA-2 source lies within the boundary of a bright-rimmed cloud,
externally illuminated by the O-stars in the W5-E cluster. This source appears to be cold, a conclusion which is confirmed by our analysis below.

\section{Properties of NGC 2559}\label{sec:ngc2559}

\subsection{Ancillary data}

We complement the SASSy observations of NGC 2559 with available archival mid- and far-IR data.
NCG2559 has been observed by both {\it IRAS\/} and {\it Akari}, providing spectral coverage from 9--160\,$\mu$m.
These data can be used to constrain the spectral energy distribution (SED) of the source.

We estimate {\it IRAS\/} fluxes using the {\it IRAS\/} Scan Processing and Integration Tool ({\sc SCANPI}
\footnote{Available at http://scanpiops.ipac.caltech.edu/9000/\-applications/Scanpi/index.html}).
Standard reduction parameters were used, taking into account the observed size of the galaxy
(${\sim}\,3\,$arcmin, \citealt{Prugniel98}; \citealt{Jarrett03}) as an additional constraint.
Although NGC 2559 is actually detected as a point source in all {\it IRAS\/} bands, we assume
a minimum distance of $6\,$arcmin from the nominal position of the source for
background subtraction. We obtain {\it Akari\/} fluxes at 9 and 18\,$\mu$m from the
{\it Akari}/IRC All-Sky Survey Point Source Catalogue \citep{Ishihara10},
and at 65, 90 and 140\,$\mu$m from the {\it Akari}/FIS All-Sky Survey Bright Source
Catalogue \citep{Yamamura10}. We decide not to use the 160\,$\mu$m measurement
owing to the extremely high noise affecting this band: this is confirmed by the $\chi^2$
increasing by about a factor 2 when this measurement is included in our fits.
In addition to that, we estimate an upper limit on the 450\,\micron\,flux from
the shorter wavelength SCUBA-2 map.
A 20 per cent uncorrelated uncertainty is applied to all data. The observed mid- to far-IR
data of NGC 2559 are shown in Table~\ref{ngc2559sed}.

\begin{table}
\label{ngc2559sed}
\caption{The observed SED of NGC 2559. Flux densities and respective errors are in Jy.
Quoted errors include an additional 20 per cent calibration uncertainty.}
\centering
\begin{tabular}{c c c c}
\hline
$\lambda$ ($\mu$m) & $S_\nu$ & $\delta S_\nu$ & Instrument \\
\hline
9 & 1.5 & 0.3 & {\it Akari}/IRC \\
12 & 1.4 & 0.3 & {\it IRAS} \\
18 & 1.8 & 0.4 & {\it Akari}/IRC \\
25 & 2.8 & 0.6 & {\it IRAS} \\
60 & 26 & 5 & {\it IRAS} \\
65 & 23 & 5 & {\it Akari}/FIS \\
90 & 39 & 8 & {\it Akari}/FIS \\
100 & 66 & 13 & {\it IRAS} \\
140 & 52 & 12 & {\it Akari}/FIS \\
450 & 2.3 & 2.3 & SCUBA-2 (u.l.) \\
850 & 0.156 & 0.04 & SCUBA-2 \\
\hline
\end{tabular}
\end{table}

\subsection{SED fitting: dust models}

We fit the dust models of \citet{DraineLi07}
\footnote{Available at http://www.astro.princeton.edu/$\sim$draine/dust/\-dust.html .}
to the compilation of available data. These models provide the dust emissivity
per hydrogen atom, $j_{\nu} (q_{\rm PAH}, U_{\rm min}, U_{\rm max})$, which is a function of
three parameters: the fraction of dust mass in PAHs, $q_{\rm PAH}$; the intensity of
the radiation field from stars heating the interstellar medium, $U_{\rm min}$; and
the intensity of the radiation field in photodissociation regions (PDRs), $U_{\rm max}$.

\begin{figure}
\label{NGC2559_fit}
\includegraphics[width=8.5cm]{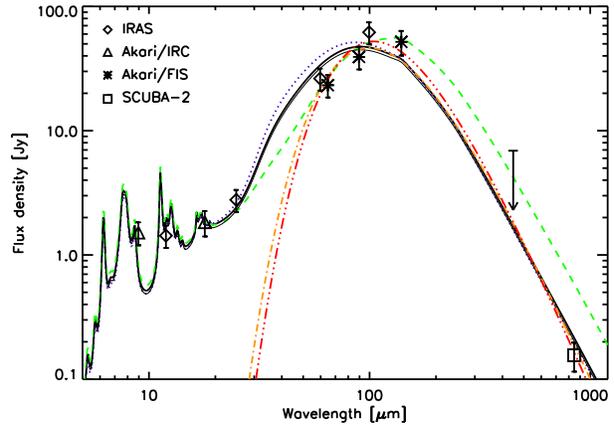}
\caption{The observed far-IR and sub-mm SED of NGC\,2559, together with dust model fits.
An arrow marks the $3\sigma$\,upper limit at 450\,$\mu$m.
The solid (black) line shows the best fit to all available data points.
The dotted (blue) line is the best fit to the dataset excluding the {\it Akari}/FIS data.
The thin solid (black) line is the best fit to the dataset excluding the {\it IRAS\/} data.
The dashed (green) line is the best fit model without the SCUBA-2 points.
The dot-dashed (orange) line is the modified blackbody fit with $\beta = 2$.
The triple-dot-dashed (red) line is the modified blackbody fit with free $\beta$.
Error bars include an uncorrelated 20 per cent uncertainty.}
\end{figure}

We apply the same general method explained in \citet{DraineLi07}, but we use only
the seven Milky Way dust model sets, as in \citet{Wiebe09}. Accordingly, we also
set $U_{\rm max} = 10^6$ (we tested the reliability of this assumption by leaving
$U_{\rm max}$ as a free parameter and found that the fit returns the same value).
We thus fit to the observed SED a linear combination of
diffuse ISM models (with $U_{\rm max} = U_{\rm min}$) and PDR models:

\begin{eqnarray*}
F_{\nu}(q_{\rm PAH}, U_{\rm min}, U_{\rm max}) & \propto & \frac{M_{\rm d}}{m_{\rm H} D^2} \\
& & \times \left[(1-\gamma)j_{\nu}(q_{\rm PAH}, U_{\rm min}, U_{\rm min})\right. \\
& & + \left.\gamma j_{\nu}(q_{\rm PAH}, U_{\rm min}, U_{\rm max}) \right],
\end{eqnarray*}
where $M_{\rm d}$ is the dust mass, $m_{\rm H}$ is the mass of a hydrogen atom and $D= 20.8\,{\rm Mpc}$
is the distance to the galaxy. We then use the derived values to evaluate the dust-weighted starlight
intensity $\langle U \rangle$.
The best-fit values for the parameters are found through $\chi^2$ minimization.
Fig.~\ref{NGC2559_fit} shows the observed SED and best-fit curves.

We notice that the {\it Akari}/FIS data are systematically low with respect
to the {\it IRAS\/} data. We thus evaluate the outcome of the fit using either the full
data set or a subset to assess the effect of this offset.
Table~\ref{NGC2559_fitval} summarizes the different fits and results.

Fitting the dust models to the full set of data points available yields a
reasonable fit, although the $\chi^2$ is relatively high. We obtain the values:
$q_{\rm PAH} = 3.2$; $\gamma = 0.01$; $U_{\rm min} = 20$; and $\langle U \rangle = 21.9$.
We derive a dust mass of $1.5 \times 10^7\,{\rm M}_{\odot}$ and a far-IR luminosity of
$2.0 \times 10^{10}\,{\rm L}_{\odot}$, from which we calculate a star-formation rate of
$2.8\,{\rm M}_{\odot}\,{\rm yr}^{-1}$ using the relation of \citet{Bell03}.
The derived dust mass is about 2 times larger than the value obtained by \citet{Bettoni03},
who find a value of $8.3 \times 10^6\,{\rm M}_{\odot}$ from data at 60 and 100\,$\mu$m.

Note that the SFR can also be estimated using the 20\,cm radio flux density along with the
FIR/radio correlation (e.g.\ \citealt{Condon92,Cram98,Hopkins01}).  With $S_{1.4}=260\,$mJy
this gives an SFR of around $10\,{\rm M}_{\odot}\,{\rm yr}^{-1}$, which is fairly consistent
with the results obtained from the FIR fit.

Exclusion of the {\it Akari}/FIS data points yields again a fit
which correctly matches the SCUBA-2 850\,$\mu$m point. 
Best-fit parameters are rather similar, the only noticeable change being the increase
of $U_{\rm min} = 25$ and of $\langle U \rangle = 27.4$, with the resulting
dust mass, FIR luminosity and SFR fairly consistent with the previous fit.

Removing the {\it IRAS\/} points from the data set yields a similar fit
with a $\chi^2$ of 10.7 with 3 degrees of freedom.  The fit parameters
are almost undistinguishable from the first fit at $\lambda \ge 30\,\mu$m.

For comparison, we also fit the models after excluding the SCUBA-2 point and upper limit.
We obtain lower values of $U_{\rm min} = 5$ and $\langle U \rangle = 5.6$).
But now the best-fit misses the 850\,$\mu$m flux density by about a factor 4.

This shows the strong leverage of the SCUBA-2 data to properly constrain the FIR and sub-mm
SEDs of galaxies. In particular, in addition to the dust mass, the value of
the starlight intensity $U_{\rm min}$ (and consequently $\langle U \rangle$) is
strongly sensitive to the 850\,$\mu$m flux density.

\begin{table*}
\label{NGC2559_fitval}
\caption{\citet{DraineLi07} model parameters for NGC 2559 and derived physical quantities.}
\centering
\begin{tabular}{c c c c c c c c c c}
\hline
Data removed & $\chi^2$ & degs. of & $q_{\rm PAH}$ & $U_{\rm min}$ & $\langle U \rangle$ & $\gamma$ &
 $M_{\rm d}$ & $L_{\rm FIR}$ & SFR \\
 &  & freedom &  &  &  &  &
  ($10^7\,{\rm M}_{\odot}$) & ($10^{10}\,{\rm L}_{\odot}$) & (${\rm M}_{\odot}\,{\rm yr}^{-1}$) \\
\hline
None & 17.7 & 7 & 3.2 & 20 & 21.9 & 0.01 & 1.5 & 2.0 & 2.8 \\
{\it Akari}/FIS & 10.8 & 4 & 3.2 & 25 & 27.4 & 0.01 & 1.3 & 2.1 & 3.0 \\
{\it IRAS} & 10.7 & 3 & 4.6 & 20 & 22.0 & 0.01 & 1.5 & 1.9 & 2.7 \\
SCUBA-2 & 3.5 & 5 & 4.6 & 5 & 5.6 & 0.01 & 5.7 & 2.2 & 3.1 \\
\hline
\end{tabular}
\end{table*}

\begin{table*}
\label{NGC2559_fitmbb}
\caption{Modified blackbody fits for NGC 2559 and derived physical quantities.}
\centering
\begin{tabular}{c c c c c c c c}
\hline
Fit & $\chi^2$ & degs. of & $\beta$ & $T_{\rm d}$ & $M_{\rm d}$& $L_{\rm FIR}$ & SFR \\
 &   & freedom & & (K) & ($10^7\,{\rm M}_{\odot}$)  & ($10^{10}\,{\rm L}_{\odot}$)
 & (${\rm M}_{\odot}\,{\rm yr}^{-1}$) \\
\hline
Fixed $\beta$ & 6.0 & 5 & 2 & 29 & 0.31 & 1.8 & 2.6 \\
Free $\beta$ & 4.7 & 4 & 2.3 & 26 & 0.47 & 2.0 & 2.8 \\
\hline
\end{tabular}
\end{table*}

\subsection{SED fitting: modified blackbody}

For comparison with the more detailed \citet{DraineLi07} models,
we also fit a modified blackbody spectrum to the observed data.
The fit is carried out assuming a shape for the modified blackbody described by the expression

\begin{equation}
S_{\nu} = \frac{M_{\rm d}\kappa}{D^2}\left(\frac{\nu}{\nu_0}\right)^{\beta}B_{\nu}(T),
\end{equation}
where $\nu_0 = 1.2\,{\rm THz} = c /(250\,\mu{\rm m})$, $\kappa$ is the
dust mass absorption coefficient at $\nu_0$, $\beta$ is
the dust emissivity index and $M_{\rm d}$ is the dust mass. Once again, the best-fit
parameters are found by $\chi^2$ minimization. We assume a mean value of
$\kappa = 0.29$ (see e.g. \citealt{Wiebe09}, although there is considerably uncertainty in
this value) and fix $\beta = 2$.

This modified blackbody fit yields consistent values of $M_{\rm d}$, $L_{\rm FIR}$ and SFR
with respect to the detailed dust models.
Table~\ref{NGC2559_fitmbb} summarizes the modified blackbody results, and the fitted curves
are shown in Fig.~\ref{NGC2559_fit}.
We see that the dust temperature of NGC 2559 is around 26--$29\,$K.  This is warmer than for most of
the galaxies studied in the SINGS sample \citet{SINGS}, as well as those studied using BLAST \citet{Wiebe09}.
This is consistent with requiring a relatively higher value of $U$ than for those other galaxies.

\section{Source Properties in W5-E}\label{sec:w5}

From our catalogue of compact sources within the W5-E region we
select two example sources, representative of two extreme regimes:
(i) the brightest source in the field, AFGL\,4029 (\citealt{Deharveng} and references therein);
and (ii) a fainter
source to the NW of the first, as an example of a potentially colder core
(labelled `cold core' in Fig.\ \ref{w5colsnr}). The SCUBA-2 flux densities
for both sources were obtained from photometry within an aperture of 30-arcsec diameter.

\begin{figure*}
  \centering
  \includegraphics[width=6.5in]{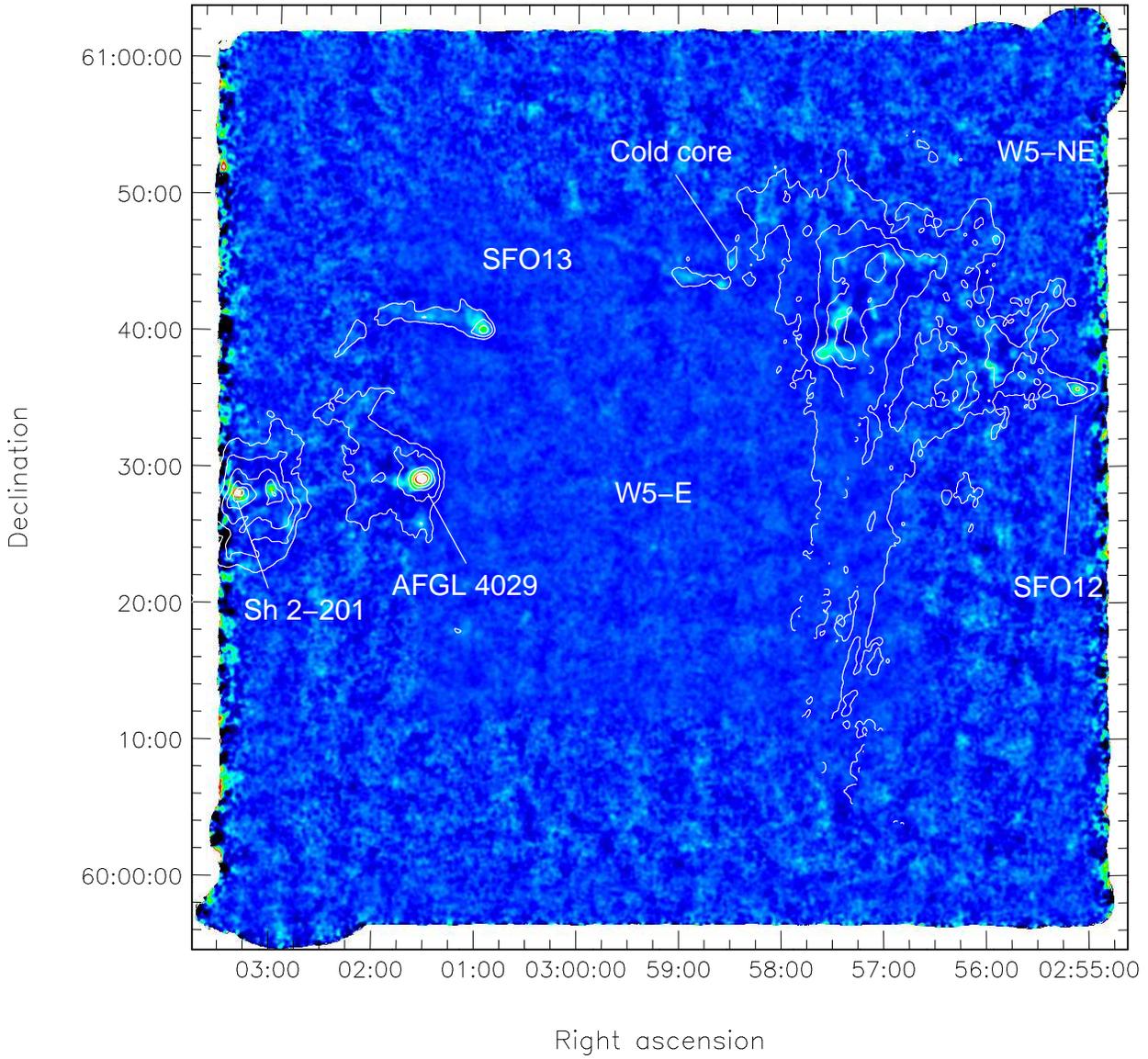}
  \caption{SCUBA-2 $850\,\mu$m emission in colour with {\it Spitzer\/} $70\,\mu$m contours overlaid.
Sources are labelled as in Fig.\,\ref{w5colsm2}.
The {\it Spitzer} contours do not close to the south of W5-NE due to a lack of coverage there.}\label{w5annotated}
\end{figure*}

Ancillary data available for this region include {\it Spitzer\/} MIPS $70\,\mu$m observations
of the whole field, as well as {\it IRAS\/}, {\it Akari} coverage, and SCUBA imaging for part
of the field.  The SCUBA-2 data are in
excellent agreement with the SCUBA measurements, being consistent within errors, but
covering a much wider area.  We show {\it Spitzer\/} $70\mu$m contours on top of the SCUBA-2 image
in Fig.~\ref{w5annotated}.

The bright source AFGL\,4029 is detected in almost all wavebands, having upper limits only
at 65 and 90\,$\mu$m. We fit a modified blackbody spectrum to the observed SED, after adding
a 20 per cent uncorrelated uncertainty to the errors, as we did for NGC\,2559.
Leaving $\beta$ as a free parameter in the fit yields a value of $\beta = 2.02$, which suggests that
$\beta = 2$ is a good assumption for the dust emissivity index. We derive a temperature $T = 31$\,K
and a dust mass of 60\,${\rm M}_{\odot}$, with a FIR luminosity of 5000\,${\rm L}_{\odot}$.  This is in reasonable
agreement with the value of 3200\,${\rm L}_{\odot}$ estimated by \citet{morgan}.

The faint source is detected only at 70 and 850\,$\mu$m. A relatively nearby 90\,$\mu$m source is detected by
{\it Akari\/}, however its position is offset by $28\,$arcsec from the {\it Spitzer\/} position.
Although this lies within the PSF of {\it Akari\/} at 90\,$\mu$m, we prefer not to use this measurement in our analysis
to avoid contamination by other potential sources (and in any case it adds little to the 70\,$\mu$m constraint).
Reliable flux densities were obtained from the {\it Spitzer\/} 24 and
70\,$\mu$m images, using the same 30-arcsec aperture as for the SCUBA-2 data. While the cold core is also
evident in the 160\,$\mu$m MIPS image, the presence of significant artefacts prevents the extraction
of a useful flux measurement.

\begin{figure*}
\label{W5E_fit}
\includegraphics[width=8cm]{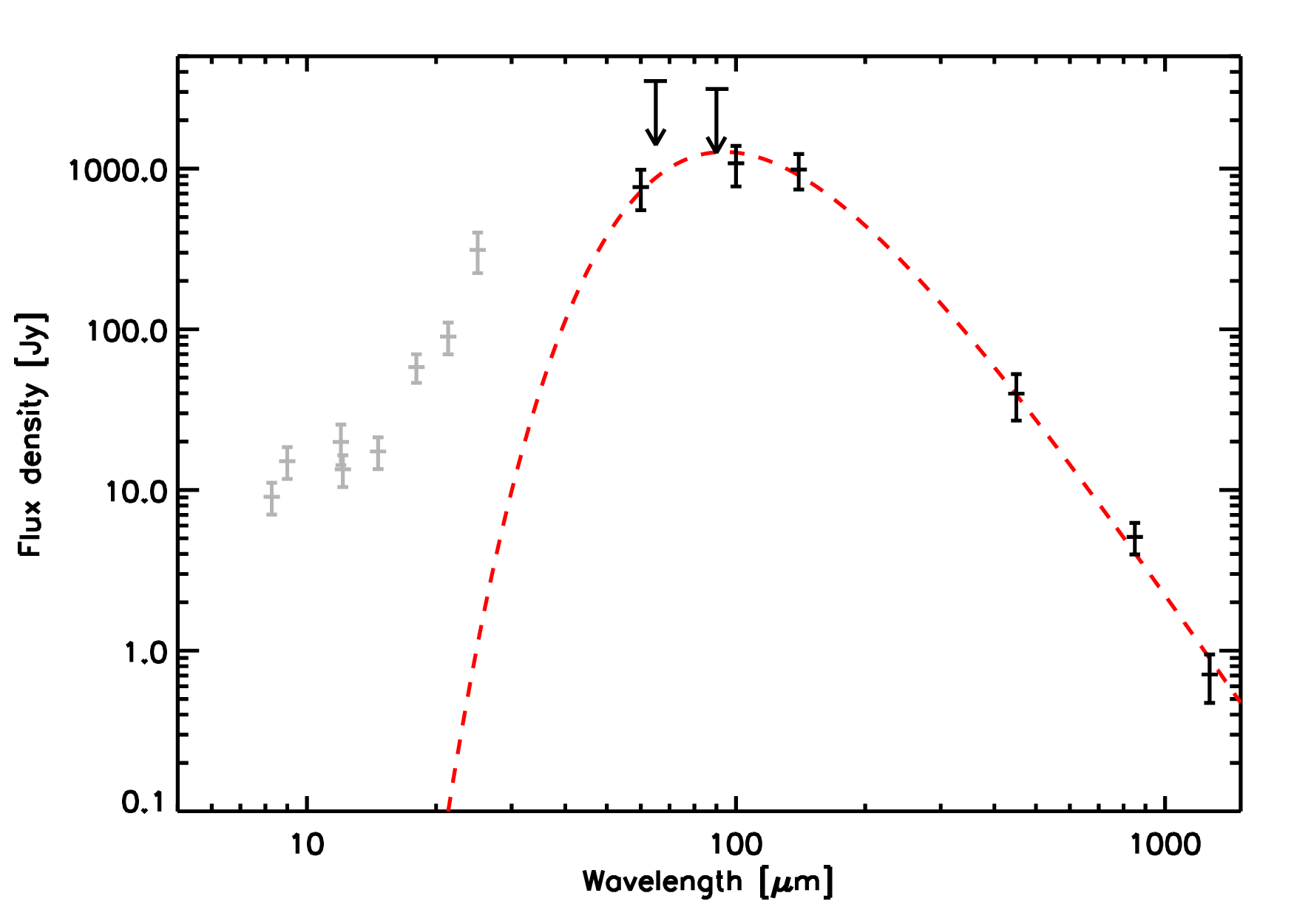}
\includegraphics[width=8cm]{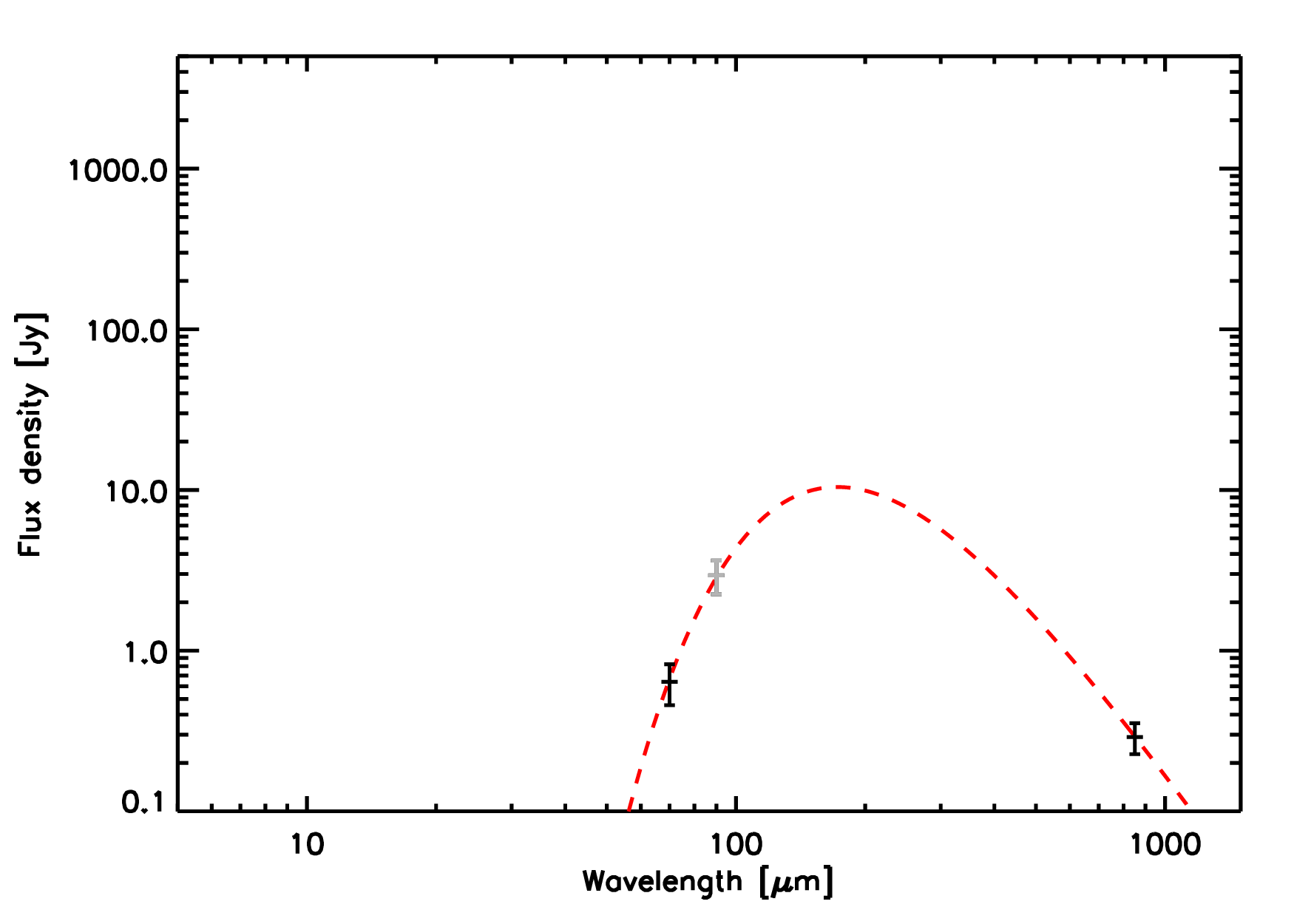}
\caption{The observed far-IR and sub-mm SED of AFGL\,4029 and of the cold source in W5-E.
Black points are used for the fit, arrows mark upper limits.
The dotted (red) line is the modified blackbody fit with $\beta = 2$.}
\end{figure*}

Although only two photometry points are available, a fit to the SED using again a modified blackbody with $\beta = 2$
yields reasonable quantities and confirms this source to be a cold protostellar core. We obtain a temperature
$T = 17$\,K, FIR luminosity of 27\,${\rm L}_{\odot}$ and a dust mass of 11\,${\rm M}_{\odot}$. 
Comparing the {\it Akari\/} 90\,$\mu$m point to the fit shows that the measurement is in excellent agreement
with the expected flux density, suggesting that this might actually be the counterpart to the same source.
The result of the fit for both sources is shown in Fig. \ref{W5E_fit}.

The readiness with which we were able to pick out a relatively cold source, even in this small pilot study,
shows that SASSy will be able to detect single low-mass cold clumps inside larger star-forming regions.
Detailed follow-up of such sources will determine where they lie on the star-formation sequence.

\section{Discussion}\label{sec:discussion}

With the current SASSy S2SRO observations, we have shown that we will be able to 
detect and catalogue real sources, both Galactic and extragalactic. Based on early commissioning estimates,
SCUBA-2 mapping during S2SRO was expected to be approximately 50 times faster than for SCUBA. A direct
comparison with the SCUBA data in the W5-E field shows the mapping speed improvement factor to be 60,
after scaling the map sizes and noise levels. Assuming similar weather conditions and a similar efficiency for the new
science-grade arrays currently being commissioned on SCUBA-2, SASSy should be able to 
map the sky at a rate of ${>}\,0.8\,{\rm deg}^2$ per hour to the target sensitivity, exceeding the capabilities of SCUBA by hundreds of times.  
However, the performance of the new $850\,\mu$m arrays are currently unknown, since they are yet to be tested on the sky,
and hence the area which will be ultimately mappable by SASSy is still uncertain.
At the present time SCUBA-2 has all eight arrays installed and is undergoing the first stages of full commissioning. It is anticipated that this will be completed in mid-2011 and the instrument made available to the community as soon as possible thereafter.  SASSy will therefore start in earnest some time in 2011.
 
Nevertheless, these initial data have shown that it is possible to reach a $1\sigma$ sensitivity to
point sources of ${\sim}\,30\,$mJy while scanning rapidly with SCUBA-2 in relatively mediocre weather conditions.
The map-making and source extraction procedures are fast and require little in the way of human intervention.
Experience with this pilot programme has already been fed back directly into the software pipeline and most of the data processing and analysis is now automated.

The science case for SASSy remains strong and has been made stronger by recent {\it Herschel\/}
discoveries.   There is still
a pressing need for a wide-area shallow $850\,\mu$m survey in the era of {\it Herschel}, {\it Planck\/} and ALMA.
This pilot study has shown that it is feasible to find sources in the shallow maps which SCUBA-2 will soon
produce routinely.

\section*{Acknowledgments}

The James Clerk Maxwell Telescope is operated by The Joint Astronomy Centre on behalf of the Science and
Technology Facilities Council of the United Kingdom, the Netherlands Organisation for Scientific Research,
and the National Research Council of Canada.  Data for this paper were taken as part of the S2SRO programme,
with Project ID M09BI142.  This research used the facilities of the Canadian Astronomy Data Centre operated
by the National Research Council of Canada with the support of the Canadian Space Agency.
This research was supported by the Canadian Natural Sciences and Engineering Research Council and enabled
through funding from the Canada Foundation for Innovation and through the CANFAR Programme, funded by
CANARIE, Canada's advanced Internet organization.
This work is based in part on observations made with the {\it Spitzer Space Telescope},
which is operated by the Jet Propulsion Laboratory, California Institute of Technology, under a contract with NASA.
This research is based in part on observations with {\it Akari}, a JAXA project with the participation of ESA.
This research has made use of the NASA/IPAC Extragalactic Database (NED) which is operated by the Jet Propulsion
Laboratory, California Institute of Technology, under contract with NASA.
The Digitized Sky Surveys were produced at the Space Telescope Science
Institute under U.S. Government grant NAG W-2166. The images of these surveys
are based on photographic data obtained using the Oschin Schmidt Telescope on
Palomar Mountain and the UK Schmidt Telescope.
The VLA is part of the US National Radio Astronomy Observatory,
a facility of the National Science Foundation operated under cooperative agreement by Associated Universities, Inc.

\bibliography{references}


\end{document}